(Research Field: Analytical and Inorganic)

*In situ* ore formation experiment: Amino acids and amino sugars trapped in artificial chimneys on deep-sea hydrothermal systems at Suiyo Seamount, Izu-Bonin Arc, Pacific Ocean

Yoshinori Takano*, Katsumi Marumo, Toshiomi Ebashi, Lallan P. Gupta, Hodaka Kawahata, Kensei Kobayashi[1], Akihiko Yamagishi[2], and Tomohiko Kuwabara[3]

*Institute of Geology and Geoinformation (IGG),
  National Institute of Advanced Industrial Science and Technology (AIST),
  AIST Central 7, Higashi 1-1-1, Tsukuba, Ibaraki 305-8567, Japan

1 Department of Chemistry and Biotechnology, Yokohama National University,
  79-5 Hodogaya-ku, Yokohama 240-8501, Japan

2 Department of Molecular Biology, Tokyo University of Pharmacy and Life Science
  1432-1 Horinouchi, Hachiohji, Tokyo, 192-0392, Japan

3 Institute of Biological Science, University of Tsukuba,
  1-1-1 Tennodai, Tsukuba, Ibaraki 305-8572, Japan

---

E-mail: takano.yoshinori@aist.go.jp

**Abstract:** The present study reports on the bio-organic composition of a deep-sea venting hydrothermal system originating from arc volcanism; the origin of the particulates in hydrothermal fluids from the Suiyo Seamount in the southern Izu-Bonin (Ogasawara) Arc is discussed with regard to amino compounds. Chimney samples on deep-sea hydrothermal systems and core samples at Suiyo Seamount were determined for amino acids, and occasionally amino sugars. Two types of chimney samples were obtained from active hydrothermal systems by submersible vehicles: one was natural chimney (NC) on a hydrothermal natural vent; the other was artificial chimneys (AC), mainly formed by the growth and deposition of sulfide-rich particulate components in a Kuwabara-type *in situ* incubator (KI incubator). Total hydrolyzed amino acids (THAA) and hydrolyzed hexosamines (HA) in AC ranged from 10.7 nmol/g to 64.0 nmol/g and from 0 nmol/g to 8.1 nmol/g, respectively, while THAA in hydrothermally altered core samples ranged from 26.0 nmol/g to 107.4 nmol/g-rock. The THAA was thus of the same order of magnitude in AC and core samples. The representative biochemical indicator ratios of β-alanine/aspartic acid and γ-aminobutyric acid/glutamic acid were low. This is consistent with a large microbial population and labile subterranean biogenic organic compounds. A micro energy dispersive X-ray spectrometer (μEDX) was used to show that the formation of massive pyrite- and chalcopyrite-ores, the major matrixes of the natural chimney structures, were processed during hydrothermal activity. These data suggest that labile particulate organic matter (POM) and/or dissolved organic matter (DOM) from hydrothermal fluid were continuously trapped to form concentrated organic matter in a sulfide-rich matrix.

Since the end of the 1970's, when high-temperature fluid venting from the seafloor was discovered at the spreading axes of the Galapagos Ridge,[1] a number of studies have revealed that hydrothermal activity is a ubiquitous phenomenon in tectonically active regions where shallow magma is present, such as mid-ocean ridges, back-arc basins, and hotspot volcanism. The peculiar phenomena of these environments have been studied from many scientific perspectives, such as geology, biology, chemistry, and physics.[2] Over the past decade, great progress has been made in our understanding of the chemistry of deep-sea hydrothermal systems. This has largely been driven by the results of field [3-5] and laboratory [2,6-8] studies exploiting recent advances in analytical geochemistry. With regard to chemical evolution and the origins of life, deep-sea hydrothermal systems are attractive environments because they represent reducing, energy-rich, and metal ion-rich conditions.[9] Anomalous concentrations of glycine in the Red Sea may also support the significance of deep-sea chemical evolution and biogeochemistry.[10] In fact, a number of submarine ecological colonies have been recognized in near black or clear smokers and the associated organic-rich seafloor mats.[11-13] Submarine hydrothermal environments support unique biologic communities based largely on biomass generated by chemolithoautotrophic microorganisms.[14] The conversion of geochemical energy to biomass places critical constraints on the biogenic productivity and community structure of vent ecosystems.[15,16]

We focus on the influence of deep-sea hydrothermal activities on biological communities of a seafloor hydrothermal system developed atop of the Suiyo Seamount in the Izu-Ogasawara island-arc in the Western Pacific. Hydrothermal systems have prevailed throughout the geological history of Earth, and ancient Archaean hydrothermal deposits may provide clues in our understanding of the Earth's earliest biosphere.[17]

Modern hydrothermal systems support a plethora of micro-organisms and macro-organisms, and provide good comparisons for paleontological interpretations of ancient hydrothermal systems. The biological environments of extreme ocean-floor vents can be well characterized by the presence of bioorganic compounds, particularly amino acids, which are common components of all organisms, and constitute a major fraction of organic matter.[18] There are, however, few reports dealing with the concentration of amino acids together with amino sugars in deep-sea hydrothermal systems. However, preliminary determinations of amino acids and the stereoisomer ratio (D/L ratio) have been reported.[19,20] The lack of evidence of abiotically synthesized amino acids, such as ω-amino acid specimens, and the unexpectedly large enantiomeric excesses of L-form amino acids support the existence of a vigorous subjacent microbial oasis, which extends the known terrestrial habitable zone.[20] Amino sugars have a wide variety of forms, ranging from minor contributions by free monomers to a dominance of polymeric forms, such as chitin, chitin peptide materials, and bacterial cell walls. Here, we report on the concentration of amino acids and amino sugars in deep-sea hydrothermal systems to compare chimney components with core samples beneath a hydrothermal vent. Two types of chimney samples were obtained from active hydrothermal systems by submersible vehicles: one was natural chimney (NC) on a hydrothermal natural vent; the other was artificial chimneys (AC), mainly formed by the growth and deposition of sulfide-rich particulate components in a Kuwabara-type *in situ* incubator (KI incubator). The results of the present study may have significant biogeochemical implications for the growth and deposition of sulfide-rich components in the sub-vent biosphere.

**Experimental**

**Sampling location of Suiyo Seamount**: The Izu-Bonin Arc lies on the eastern rim of the Philippine Sea plate. The arc is about 1,200 km long, extending from the Izu Peninsula (35ºN, 139ºE) to Minami-Iwojima Island (24ºN, 141ºE). The arc belongs to the circum-Pacific island-arc system, and is adjacent to the Northeast Japan Arc to the north and the Mariana Arc to the south. Numerous volcanic islands and submarine volcanoes run parallel to the Izu-Bonin trench and form the volcanic front of this intra-oceanic island-arc system. The southern Izu-Bonin Arc, which is divided by the Sofugan tectonic line from the northern Izu-Bonin Arc (Fig. 1-(a)), [21] is thought to have become active at around 42 Ma.[22] The Shichiyo Seamount chain forms a volcanic front (Fig. 1-(b)) around which the arc crust is believed to become thinner than that in the northern part.[23] The Suiyo Seamount, one of the volcanoes in the Shichiyo chain, has two major peaks, located on the eastern and western sides of the seamount. Dacitic rocks of a calc-alkaline rock series and low-potassium andesites have been recovered from this area,[24] and seafloor hydrothermal alteration at Suiyo was initially reported with regard to the mineralogical characteristics, [25-27] and isotopic behaviors.[28]

As to the terrestrial origin of organics on the Suiyo Seamount, it was reported that the total fatty acid composition of surface sediments obtained from the Suiyo hydrothermal system of the Izu-Bonin Arc did not account for a significant portion of the organic matter present.[29,30] Analytical results of the surface sediments revealed very low concentrations of terrestrial sediments. In addition, an age determination of unaltered dacite by the Ar-Ar method yielded 9,000 ± 8,000 yrBP, thus suggesting zero age.[26,27] The caldera floor is predominantly covered with a sandy sediment and hydrothermal precipitation, and lacks any evidence of muddy pelagic sediment. Hydrothermal circulation reaches the region adjacent to the magma source, and volatile constituents are extracted by water-rock interaction.[17] Numerous short black smokers and clear smokers were observed on the volcanic pumice floor.

**Chimney Sampling:** Deep-sea hydrothermal chimney samples were collected as part of the Archaean Park Project during a cruise over the Suiyo Seamount (28°33'N, 140°39'E) in the Pacific Ocean in 2001 and 2002. The deep-sea subterranean biosphere and geochemical interaction were examined by taking natural chimney (NC) samples, named HY#12-CM, using submersible vehicles *Hakuyo 2000* and R/V *Shinsei-Maru*, during survey of hydrothermal areas in the caldera.[17] The location of the hydrothermally altered area in Suiyo caldera is shown in Fig. 2. The artificial chimney (AC) was created as follows: the KI incubator (Rigo Co.) was placed on an active hydrothermal vent (near APSK 07 site) at 28° 34.274'N, 140° 38.602'E using the remote observation vehicle *Hakuyo 2000* and the R/V Shinryu-Maru (SNK Ocean). The KI incubator was then recovered by the submersible vehicles *Shinkai 2000* and R/V *Natsushima* (NT02-09) after one month. The schematic experimental design and the KI incubator apparatus are shown in Figs. 3 and 4, respectively. Fresh apatite rock (produced from Hull, Quebec, Canada), dacite pumice (purchased from a local market), and dacite rock (produced from Obara, Miyagi, Japan) were placed in the incubator as carriers in order to trap and culture microorganisms in the hydrothermal fluid. Although we analyzed the deposited sulfide-rich components, we did not use fresh apatite rock, dacite pumice, and dacite rock for the chemical analysis.

**Core sampling:** Deep-sea hydrothermal core samples were collected as part of the Archaean Park Project during a cruise over the Suiyo Seamount (28°33'N, 140°39'E) in the Pacific Ocean in June, 2001. The deep-sea subterranean biosphere and geochemical interaction were examined by taking core samples using a fixed seafloor benthic multi-coring system (BMS) for pinpoint drilling.[31,32] The diameter of the core was approximately 44 mm. Recovered core samples and the mineral assemblage are shown in Fig. 5. Temperature logging of boreholes after BMS drilling was performed using a custer-type thermometer,[33] and peak temperatures were 304 °C and 156 °C at sites APSK 05 and APSK 07, respectively. The temperature of the hydrothermal fluids discharging into the BMS drill holes was measured the manned submersibles *Shinkai 2000* and *Hakuyo 2000* during ROV dives performed after BMS drilling. The temperature of the hydrothermal fluids was found to be 308.3 °C and 272.0 °C at sites APSK 05 and APSK 07, respectively.[32] The core samples from drill holes APSK 05 and APSK 07 contained hydrothermal sulfides, anhydrite, barite, chlorite / montromollonite mixed-layer minerals, mica, and chlorite with little or no feldspar and cristobalite.[26,27] Hydrothermal clay minerals changed with the depth from montmorillonite to chlorite and mica through chlorite / montmorillonite mixed-layer minerals.[26,27]

**Removal of silicates from the core samples.** All of the reagents used were of ultra-pure HPLC grade. Deionized water was further purified with a Millipore Milli-Q LaboSystem™ and a Millipore Simpli Lab-UV™ (Japan Millipore Ltd., Tokyo, Japan) in order to remove both inorganic ions and organic contaminants. All glassware was heated for 2 hours in a high-temperature oven (Yamato DR-22) at 500 °C prior to use in order to eliminate any possible organic contaminants.

The pre-treatment of core samples is described elsewhere.[20,35] In brief, approximately 1.0 g of a dry powdered sample was placed in a Teflon tube that had been cleaned by soaking in 7 M $HNO_3$ overnight and rinsed in Milli-Q-water (Millipore Corp.). Next, 10 ml of a 5 M HF-0.1 M HCl mixture was poured into the Teflon tube, which was placed in a metal holder. This was continuously heated at 110 °C for 16 hours in order to extract organics from the silicate matrix. After HF-HCl degradation, Teflon tubes were placed on a hot plate in a draft chamber in order to evaporate acids. Organic residues were extracted using pure water with ultra-sonication. The aqueous fraction was filtered with a GF/A 1.6-μm glass filter, and was then freeze-dried in a glass test tube. A total of 2 ml of 6 M HCl was added to each test tube in order to obtain the total hydrolyzed amino acid fraction (THAA). Test tubes were sealed and placed in a block heater, and were heated for 2 hours at 110 °C. The hydrolyzed portion containing amino acids and amino sugars was cooled to room temperature, and then filtered through a membrane filter (pore size, 0.45 μm) using a disposable syringe. The ampoule was washed twice with Milli-Q water, which was then passed through the same filter. The hydrolysates were then dried *in vacuo* using a diaphragm pump.

After drying, the portions were adjusted to pH 1 with 0.1 M HCl, followed by desalination with a AG-50W-X8 (200-400 mesh) cation-exchange resin column (Bio-Rad Lab.). Before applying the sample to the column, the resin was cleaned by passing 1 M HCl, $H_2O$, 1 M NaOH, and $H_2O$ successively through the column. Immediately prior to applying the sample, the resin was reactivated with 10 ml of 1 M HCl and rinsed with 10 ml of $H_2O$. The amino acid fraction was eluted with 10 ml of 10% $NH_3$. The eluate was freeze-dried and redissolved in 1.0 ml of 0.1 M HCl before injection into a liquid-chromatographic system. In order to prepare an unhydrolyzed fraction of amino compounds, about 2 g for each core sample was extracted with 3 ml of water by shaking for 1 h. The extraction was repeated three times, and the extracted solutions were combined together. Then, after dryness, the fraction was analyzed in the same manner as the hydrolyzed fraction.

**Determination of amino acids and amino sugars:** The concentrations of hydrolyzed amino acids and hexosamine were determined using an ion-exchange HPLC system, which was composed of three high-performance liquid-chromatograph pumps (Shimadzu LC-9A), and a strongly acidic cation-exchange column (Shim-pack Amino-Na column, 10 cm x 6.0 mm i.d.).[36,37] The column was maintained at 60 °C during analysis and was preceded by a pre-column for reducing the effects of ammonia and other unwanted chemicals. Post-column derivatization of amino acids and amino sugars was achieved with two reaction reagents in a chemical reaction box at 60 °C: phthalaldehyde (OPA) reagent and carbonic acid-boric acid buffer solution (pH 10.0), which were supplied with peristaltic pump (Shimadzu PRR-2A). The resulting fluorescent complexes were detected and estimated using a Shimadzu RF 550 fluorescence detector (excitation wavelength 355 nm, emission wavelength 397 nm). Ten micro litters of a standard, containing 1 nmol of each amino acid and amino sugar was also analyzed after every 5th sample. The resulting peak areas were computed with a Shimadzu C-R4A Chromatopac integrator. The detection limit in this method was ca. 1 nmol/g for amino acids and amino sugars. The data reproducibility in the present study was better than ± 5% for the relative molar concentrations of the amino acids and amino sugars.

The separation of D- and L-amino acid enantiomers was achieved by high-performance liquid chromatography (RP-HPLC) using HPLC pumps (CCPM II, TOSOH), a reversed-phase column (YMC-pack Pro C18, 250 mm x 4.6 mm i.d.), and a TOSOH FS fluorometric 8020 detector (excitation wavelength 355 nm, emission wavelength 435 nm). An aliquot of a pre-treated sample was mixed well with N-AceCys and OPA in a glass vial, and injected into the HPLC column. Gradient elution was applied using the following eluents: A, 40 mM sodium acetic acid buffer (pH 6.5); B, 100% methanol (ultra-pure HPLC grade). The gradient program was performed as follows: 10 min (Eluent B: 0%) – 25 min (Eluent B: 10%) – 65 min (Eluent B: 20%) – 80 min (Eluent B: 20%) – 85 min (Eluent B: 40%) – 115 min (Eluent B: 60%) – 120 min (Eluent B: 80%) – 135 min (Eluent B: 0%).[38]

**Result and Discussion**

**Concentration of amino acids in artificial chimney (AC) and core samples:** Blank analysis of amino acids during laboratory handling gave trace levels of glycine. Nineteen types of protein and non-protein amino acids were quantitatively determined. The abundance of amino acids in the AC samples ranged from 10.7 to 64.0 nmol/g-rock, as shown in Table 1. Fresh samples of AC predominantly contained glycine, with lesser concentrations of proteinous amino acids, such as alanine, serine, and aspartic acid. Fig. 6 shows the molar ratios (mole%) of the amino compounds in the AC. Although non-proteinous amino acids, such as β-alanine, α-aminobutyric acid, and γ-aminobutyric acid, have been identified as major products in laboratory formation experiments simulated hydrothermal systems,[8,39] they were present as only minor constituents in the present chimney samples. Hence, biogenic labile flux of particulate organic matter derived from hydrothermal fluid was experimentally suggested. Methionine is known to be highly susceptible to hydrolytic loss during acid hydrolysis.[20] Therefore, the measured concentration of methionine in the sediments may be higher than the actual concentration.

In core samples of the APSK 05 and 07 series, we could not detect only amino compounds in the present analytical systems. Thus, amino acids and amino sugars did not occur in free, water-soluble form; strong acid hydrolysis was required for their release from organic matter and the mineral matrix. Only trace amounts of some proteinous amino acids were detected in some sequences of the sediment. This suggests that combined amino compounds, rather than free amino compounds, were the major components in organic matter. Consequently, combined-form analogs in the chimney structure might be more stable than free-form analogs.

**Concentration of amino sugars in an artificial chimney (AC) :** Hexosamines (HA), such as glucosamine (2-amino-2-deoxyglucose) and galactosamine (2-amino-2-deoxygalactose), are the major amino sugars in sedimentary organic matter. Glucosamine (Gluam) and Galactosamine (Galam) occur as structural components in a large group of substances, the mucopolysaccharides; they have been found in combination with mucopeptides and mucoproteins.[40] Amino sugars have a wide variety of forms, ranging from free monomers to the more dominant polymeric forms, such as chitin,[41] chitin peptide materials,[42] and bacterial cell walls.[43] Gluam is a particularly common micro arthropod chitin.[44,45] The origin of Galam is less clear, although Sowden and Ivarson (1974)[46] showed that little, if any, Galam was synthesized by bacteria in incubation experiments.

As shown in Table 1, the major amino component was proteinous amino acids; Gluam and Galam were not detected, or were present as very minor components. From a stereochemical perspective, Gluam and Galam are epimeric amino sugars. The ratio of Gluam/Galam was 0.78 and 0.90 for artificial (AC6) and natural chimneys, respectively. The total hydrolyzed amino acids (THAA) to the total hexosamines (HA) ratio (THAA/HA) varied from 7.9 to 132.0 for artificial (AC6) and natural chimneys (NC), respectively. Thus, amino acids are far more predominant than amino sugars in deep-sea hydrothermal systems. Different microorganisms produce different amino sugars,[47] but Gluam and Galam were very minor components in experiments aiming to characterize their origins.

**Concentration of amino compounds in natural chimney (NC):** With regard to amino compounds in NC, the detected amino acids and hexosamines were 398.1 nmol/g-rock and 3.0 nmol/g-rock, respectively. The NC also predominantly contained glycine, with lower concentrations of proteinous amino acids, such as alanine, serine, and aspartic acid, which was similar to the artificial chimney. The molar ratios (mole%) of glycine and alanine were 19.2% and 9.8%, respectively. Particulate organic matter (POM) and/or dissolved organic matter (DOM) from hydrothermal fluid might be continuously trapped to form concentrated organic matter in a sulfide rich matrix. As shown in Fig. 7, the relationship between the massive pyrite- and chalcopyrite-ores as the major matrix of the natural chimney structures was clarified by a micro Energy Dispersive X-ray spectrometer (μEDX) (Rayny EDX-1300, Shimadzu Co.).  This is suggestive of different cycles of hydrothermal activity: one high-temperature cycle dominated by pyrite formation, and a later cycle resulting in chalcopyrite formation.[48]  The mixing of hydrothermal fluid and seawater at temperatures of >150 °C causes saturation and precipitation of anhydrite,[49] which provides structural support and sites for sulfide nucleation.  The precipitation of chalcopyrite produces a sulfide conduit lining, and advection of hydrothermal fluid through the permeable wall causes a replacement of anhydrite with chalcopyrite, and overgrowth of earlier-formed sulfides.  Detailed modeling of the reaction-transport processes within chimney walls[50,51] has enhanced our understanding of the development of the mineralogical zoning of those walls.  The high concentrations of amino acids in natural chimneys imply that the concentrated matrix of organic compounds are trapped during chimney growth.

**Vertical distribution of amino acids in sub-vent system:** The concentrations of amino acids in the sub-vent core samples of APSK 07 site are summarized in Table 2. Nineteen types of proteinous and non-proteinous amino acids were quantitatively determined. Glycine was the most abundant amino acid, on average, followed by alanine. The concentration of glycine ranged between 2.9 and 31.5 nmol/g-rock.  Other protein amino acids, such as valine, aspartic acid and glutamic acid, were also among the major constituents, as shown in Fig. 7.  Thus, the relative abundance of each amino acid did not vary significantly with the depth.  The THAA content was on the order of $10^1$–$10^2$ nmol/g-rock for all samples.  The determined values of the artificial chimneys and the sub-vent core samples were on the same order of magnitude.

The concentrations of total organic carbon (TOC), and total nitrogen (TN) in the core samples were on the order of $10^1$ to $10^2$ μg C/g-rock and $10^1$ to $10^3$ μg N/g-rock at another hydrothermal area.[20] Although the C/N ratio for bacteria-inhabited hydrothermal vent environments is not known, the present values are notably rich in nitrogen. Fourier-transform infrared (FT-IR) analysis of clay minerals[25] in the core samples suggest that ammonium ions are present at interlayer sites (Fig. 8). Ammonia is present in hydrothermal fluid at concentrations ranging from 10 to 20 μM, and is an important nutritional source of nitrogen for the sub-vent ecosystem.[20] This trend is similar to the distribution of total amino acids, suggesting that the organic carbon and amino acids are derived from in situ organisms, either viable or nonviable.[20]

The vertical distribution of amino acids in these samples differs from typical ocean seafloor sediment in that the concentration of amino acids in normal, simple sedimentation decreases rapidly with depth due to diagenesis.[52] Diagenesis in sediment may cause the decomposition of amino acids *via* decarboxylation, by which aspartic acid will alter to β-alanine by specific decarboxylation of the α-carboxyl group.[53] Hence, regulated correlations between such dicarboxylic amino acids groups and ω-amino acids groups were observed in the vertical distribution of accumulated sediment, as was seen in previous studies, not only oceanic sediments,[52] but also terrestrial sediments.[54]

In the present study, however, the correlation in the relative abundance of β-alanine and aspartic acid with the depth was not observed.[20,35] In fact, the highest concentration of amino acids was observed not at the surface, but in the sub-seafloor of unconsolidated volcanic sands and pumice fragments. This may be due to fluids that migrate upward from deeper levels in the sub-vent, thus supplying energy and organic compounds. The movement of hydrothermal fluid may form veins toward the seafloor and/or black smoker chimneys. Consequently, the lack of evidence concerning abiotically synthesized amino acids of ω-amino acid specimens and the abiotic tendency of products support the existence of a vigorous subjacent microbial oasis, which extends the known terrestrial habitable zone.[20,35]

**Origins of amino acids and microbial activity:** As shown in Fig. 9, in some parts of the sub-vent column, the D/L ratios of glutamic acid and alanine were predominant over aspartic acid. In the abiotic formation of amino acids, the D/L ratio of amino acids converges at 1.0 due to the formation of a racemic mixture.[9,55] The large enantiomeric excess of L-form amino acids observed in the present study may indicate that the amino acids were derived from a biotic flux in the sub-vent biosphere. This implication becomes even stronger when considering previous reports that the racemization rate constant is higher under hot geothermal conditions.[56-58] The D/L ratio of amino acids in the present core samples as well as in hydrothermal water samples, therefore, reflects microbiological activity.[59] In addition to hydrothermal racemization, the D-amino acids may also originate from peptideglycans in bacterial cell walls.[60] Peptideglycans are a product of bacterial metabolism and the principal biochemical sources of D-amino acids. D-Ala, and D-Glu are among the most common D-amino acids found in bacterial cell walls.[61] The low D/L ratio for aspartic acid, glutamic acid and alanine in the vertical profile may have two origins of D-form amino acids: i) racemerized D-amino acids from protein L-form analogs, and ii) minor contribution of D-amino acids derived from microbial constituents, such as bacterial cell walls. We concluded that the amino acids detected in chimneys and sub-vents in the Suiyo Seamount are indicative of labile biogenic compounds, rather than an abiotic chemical synthetic origin.

Since the discovery of hyperthermophilic microbial activity in hydrothermal fluids recovered from smoker vents on the East Pacific Rise, a number of anaerobic sulfur-dependent heterotrophic hyperthermophiles have been isolated from high-temperature and high-pressure environments, such as terrestrial hot springs and submarine hydrothermal systems. Most of the marine forms belong to the domain *Archaea*, with the exception of the genus *Thermotoga*.[62] Heterotrophic organisms utilize a wide variety of organic compounds, including carbohydrates, amino acids, organic acids, and alcohols, as sole carbon and energy sources. Isolated hyperthermophilic heterotrophic archaea and the requisite amino acid assemblages have been characterized experimentally.[63] Among these, the genus *Thermococcus* has been grown on protein mixtures, casamino acids (amino acid components hydrolyzed by casein), and purified proteins (e.g. casein and glatin), but not on carbohydrates or organic acids.[63] Most of the $S_0$-dependent hyperthermophilic heterotrophs isolated from marine hydrothermal vents require a complex proteinaceous substrate.[64] Thus, microbial communities in hydrothermal areas favor not only free amino acid analogs, but also combined proteinaceous substrates. This indicates the consumption and reuse of a biological source of amino acids by subterranean microorganisms in order to form a habitable zone. Therefore, the concentration of amino acids may be closely related to the distribution of deep-sea subterranean microbial activity.

It is interesting to note that microbes at densities of approximately $10^4$–$10^5$ cells/ml were found in 308 °C hydrothermal fluid from a drill hole in the Suiyo hydrothermal area when the drill hole was encased with metal to block the infiltration of interstitial water.[65] The currently accepted thermal limit of life is 113 °C,[66] and although some proteins from hyperthermophiles are more active at high pressure (Bernhardt, 1984), high pressure does not increase the thermal stability of micromolecules. The recent discovery of a microbe living at 121 ºC has broken the established temperature limit and extended the zone of microbial habitable temperatures.[67] The hydrothermal gradient zone may be such that the optimum fluid temperatures for microbial life occur in the sub-vent habitable regions. The microbial diversity and populations in a hydrothermal plume that was present inside the caldera of the Suiyo Seamount were investigated by performing a phylogenetic analysis of 16S rRNA gene and by fluorescence in situ hybridization (FISH).[68] An indicator of turbidity, the vertical total cell count varied from $5.6 \times 10^4$ to $1.1 \times 10^5$ cells/ml. The Suiyo Seamount caldera functions as a natural continuous incubator for microbes in the deep-sea environment.[68,69] Not only the sub-vent environment, but also the hydrothermal plume, are habitable for microbes.

**Conclusions**

The present study shows the following characteristics with regard to amino compounds trapped during the *in situ* formation of artificial chimneys (AC) on deep-sea hydrothermal systems:

1) Chimney samples were obtained from active hydrothermal systems by submersible vehicles; the chimney was mainly formed by the growth and deposition of sulfide-rich particulate components using the KI incubator. The abundance of amino acids in the AC ranged from 10.7 to 64.0 nmol/g-rock. Fresh samples of AC predominantly contained glycine, with lower concentrations of proteinous amino acids such as alanine, serine and aspartic acid. The THAA was of the same order of magnitude in both the AC and core samples. The ratios of β-alanine/aspartic acid and γ-aminobutyric acid/glutamic acid trapped in the AC were low, as were the D/L ratios in core samples, which is consistent with a large microbial population and labile subterranean biogenic organic compounds. Consequently, the amino acids detected during *in situ* formation at Suiyo Seamount appear to be of biological origin rather than the result of abiotic chemical synthesis.

2) With regard to natural vent chimneys (NC), amino acids and hexosamines were detected at 398.1 nmol/g-rock and 3.0 nmol/g-rock, respectively. The concentrations in solidified NC were far higher than those in AC. This suggests that particulate organic matter (POM) and/or dissolved organic matter (DOM) from hydrothermal fluid might be continuously trapped in order to form concentrated organic matter in a sulfide-rich matrix. Experimentally, a micro energy dispersive X-ray spectrometer (μEDX) was used to confirm that the formation of massive pyrite- and chalcopyrite-ores, the major matrixes of natural chimney structures, were processed during hydrothermal activity.

As to thermal gradient zone in chimneys, there are drastic thermal changes between the interior, middle, and exterior portions. Exterior portions are exposed to very low temperatures (ca. < 4 °C) from the ambient sea-water, while interior and middle portions are exposed to very high temperatures. We are thus required to show a correlation between microbial activity and biochemical indicators, focusing on thermal gradient evaluations in future sub-vent studies.

**Acknowledgements:** The authors would like to thank the scientific crews of the *Shinsei-Maru* and *Natsushima* for their endeavors in surveying and sampling. The authors also appreciate the operating team of the benthic multi-coring system (BMS) during the *Hakurei-Maru* II cruise. We are grateful to the chief scientist, Prof. Tetsuro Urabe, of the Department of Earth and Planetary Science at the University of Tokyo and all scientists who participated in the cruise. The authors would like to thank Ms. M. Nakashima of the Institute for Marine Resources and Environment at the National Institute for Advanced Industrial Science and Technology and Mr. T. Horiuch, Ms. Y. Edazawa, Mr. T. Kaneko of Yokohama National University for their experimental help, reviews and discussions. This research was funded by the Ministry of Education, Culture, Sports, Science and Technology (MEXT) of Japan through the Special Co-ordination Fund for the Archaean Park Project; an international research project on interaction between the sub-vent biosphere and the geo-environment. The authors express their sincere thanks to two anonymous reviewers for constructive comments which helped to improve the earlier version of the manuscript.

----

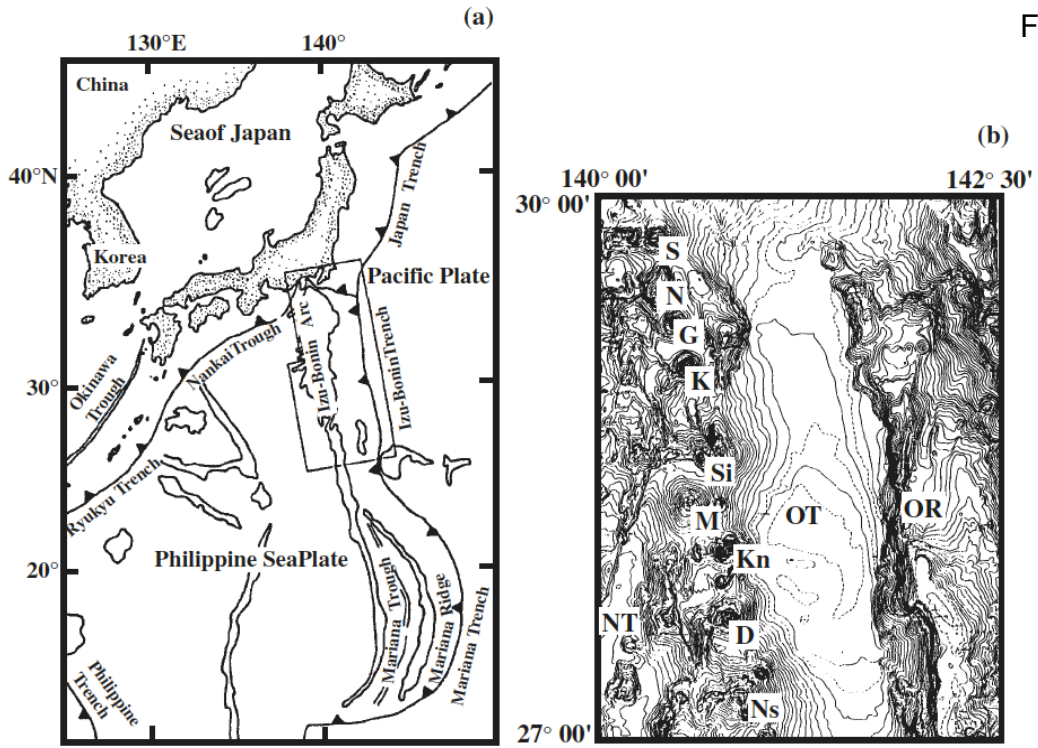

Fig. 1
(a) Geological location of the Izu-Bonin Arc on the eastern edge of the Philippine Sea plate, western Pacific ocean. (b) Topographic map of the Suiyo seamount in the Shichiyo seamount chain. Cited from the Ref. 21.

Abbreviations. *Si* = Suiyo Seamount; *OR* = Ogasawara Ridge; *OT* = Ogasawara Trough; *S* = Sofugan Island; *N* = Nichiyo Seamount; G = Getsuyo Smt.; K = Kayo Smt.; M = Mokuyo Smt.; Kn = Kinyo Smt.; D = Doyo Smt.; *Ns* = Nishinoshima Island.

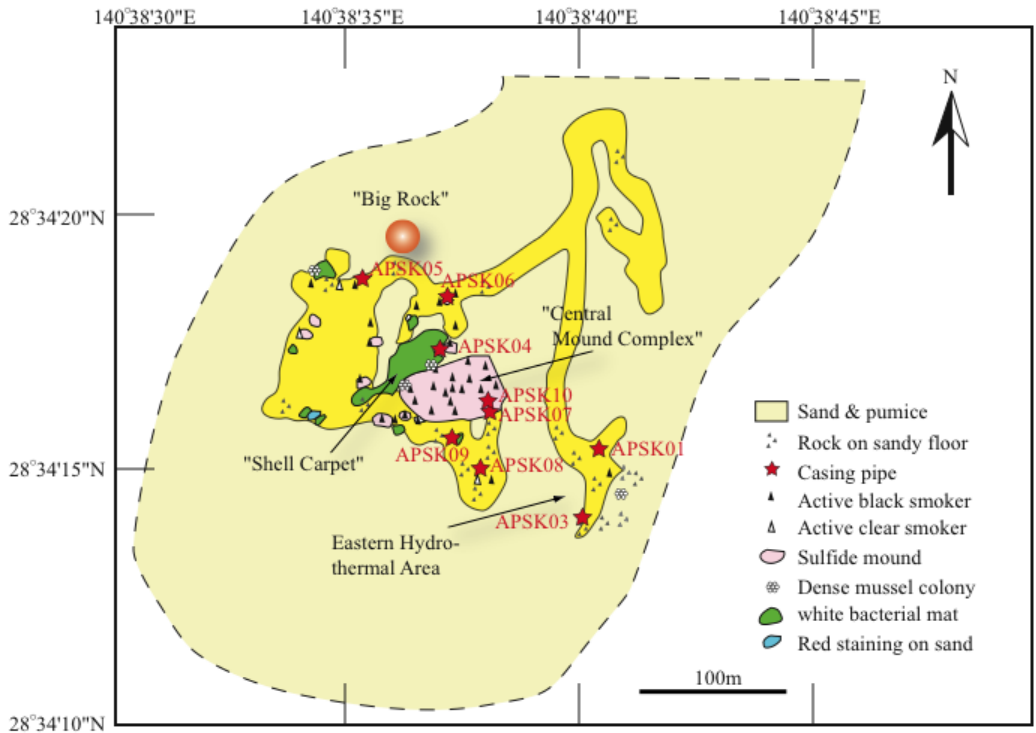

Fig. 2
Distribution of chimneys, mounds, and BMS drilling sites in the bottom of the caldera at Suiyo Seamount. Cited from the Ref. 27.

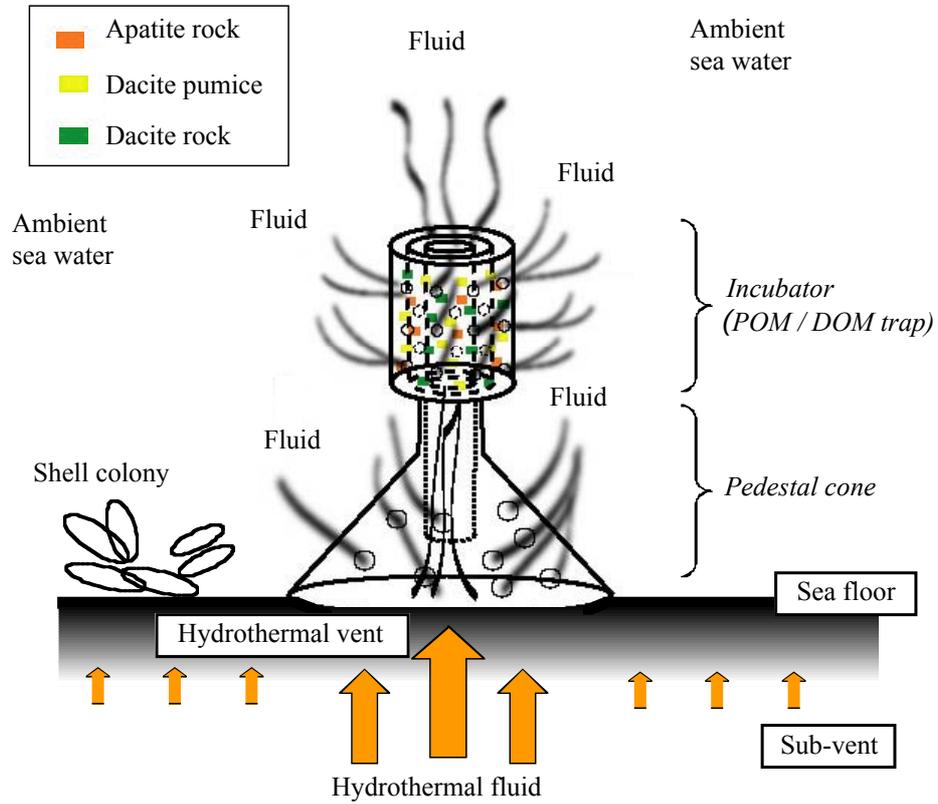

Fig. 3
Schematic design of the Kuwabara type *in situ* incubator system on deep-sea hydrothermal systems at Suiyo seamount, Izu-Bonin arc, Pacific ocean.

Figure 4

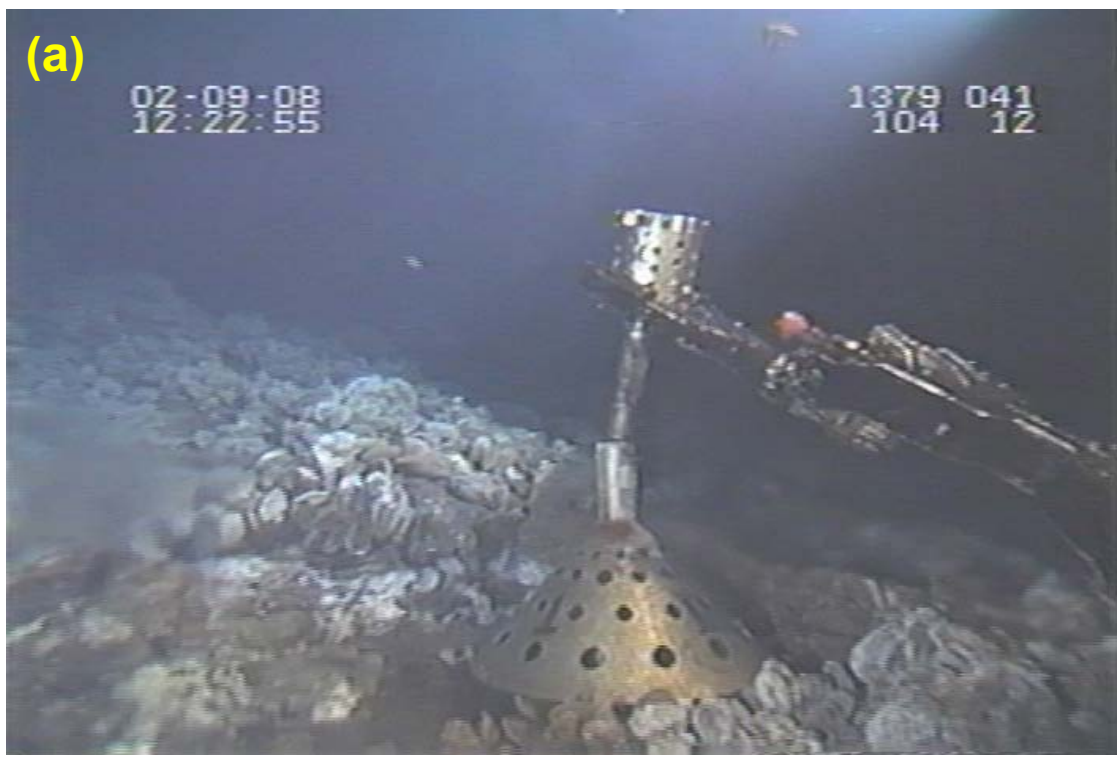

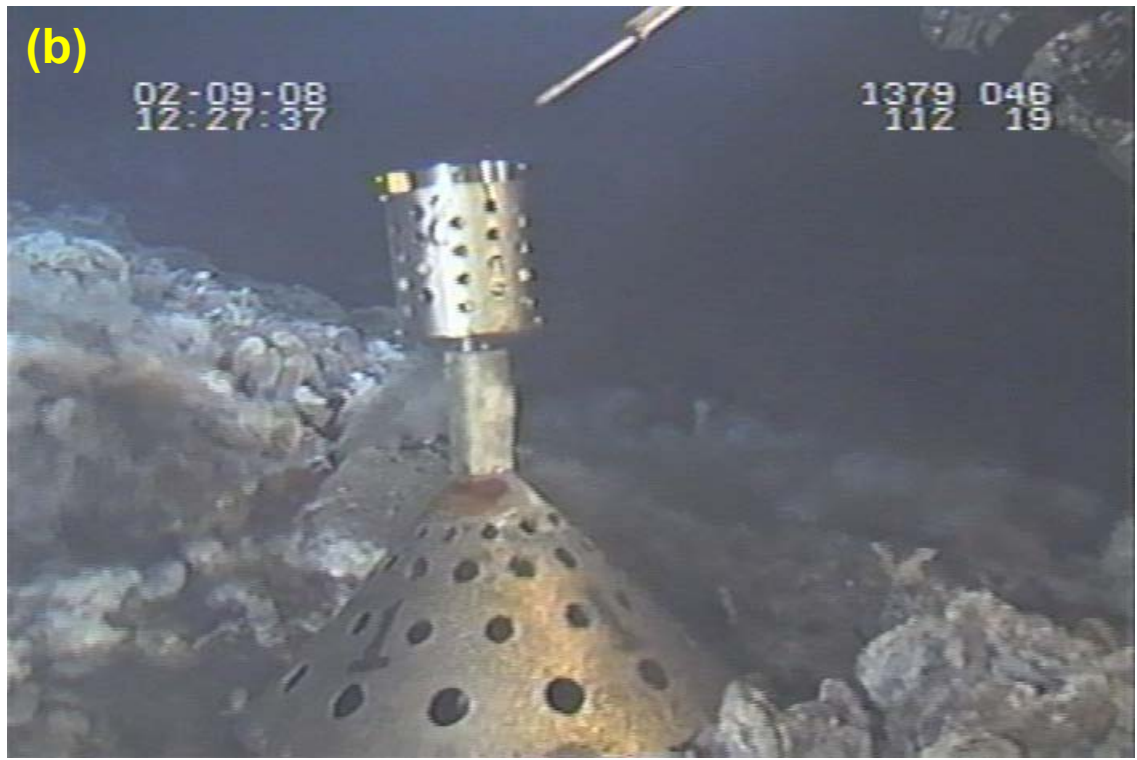

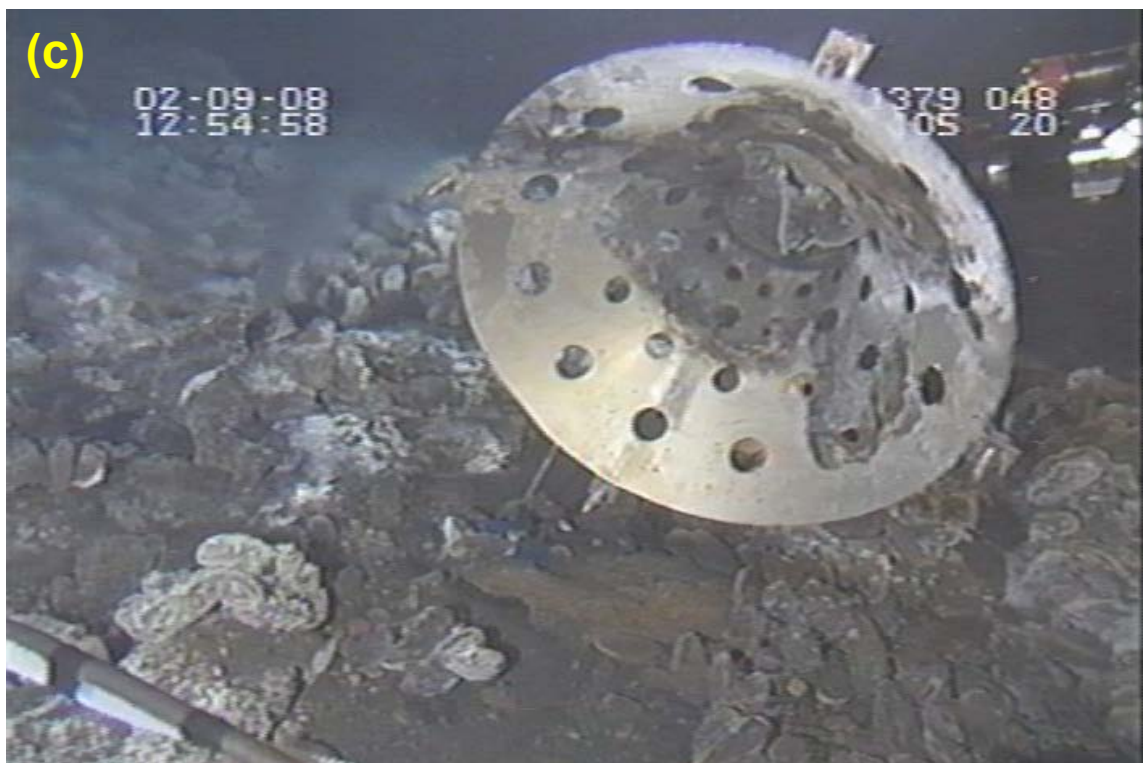

Fig. 4
Apparatus of the Kuwabara type *in situ* incubator system on the experimental site captured by *Shinkai* 2000. Each picture shows the insertion between the incubator and pedestal cone, whole appearance during deployment, and recovery of the Kuwabara type *in situ* incubator system shown in Figs. 4-(a), (b) and (c), respectively.

**(a)**

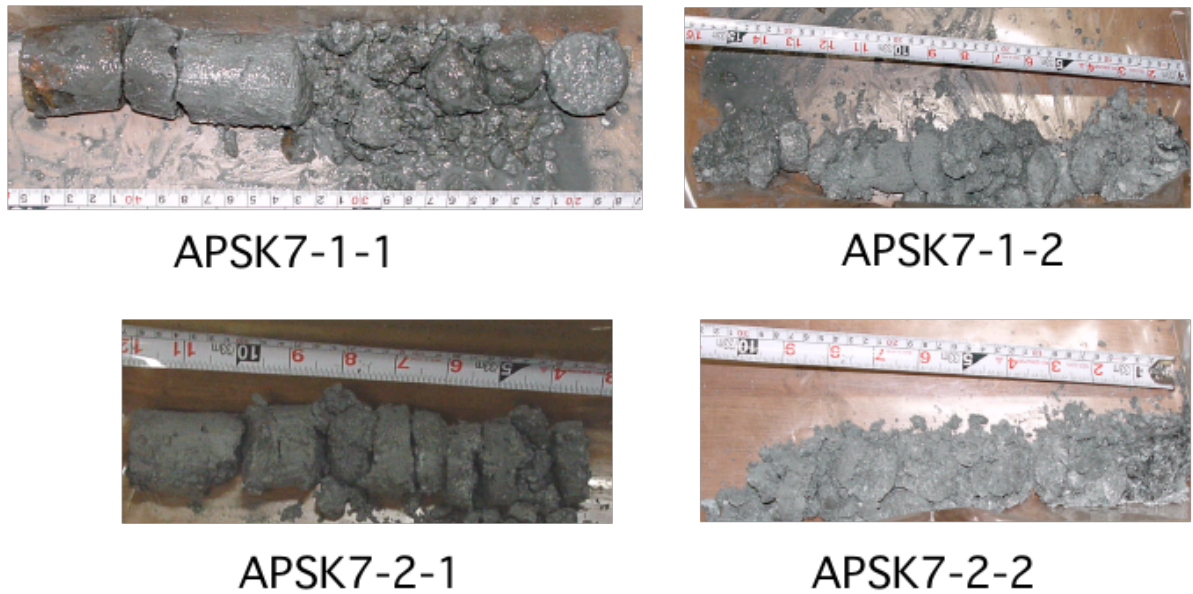

Fig. 5-(a)
Representative sub-vent core samples recovered by benthic multi-coring system (BMS) at APSK 07 site. Hydrothermal alteration was observed in terms of formation of clay analogs.

**(b)** 

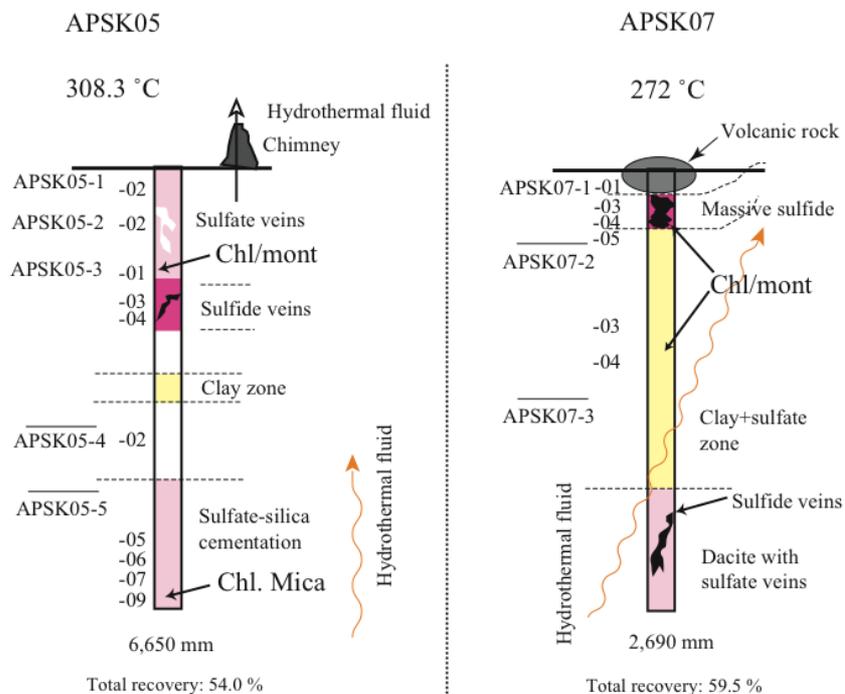

Fig. 5-(b)
The vertical variations in the mineral assemblages. The core profile is characterized by dacitic lava and/or pyroclastic rocks at the surface underlying unconsolidated volcanic sands and pumice fragments; a sheath of clay minerals and anhydrite cement with minor pyrite and other sulfide minerals that acts as a cap rock of the geothermal system; and hydrothermal fluid ponding beneath the sheath. Abbreviations. Chl., Chlorite; Mont. Montmorillinite.

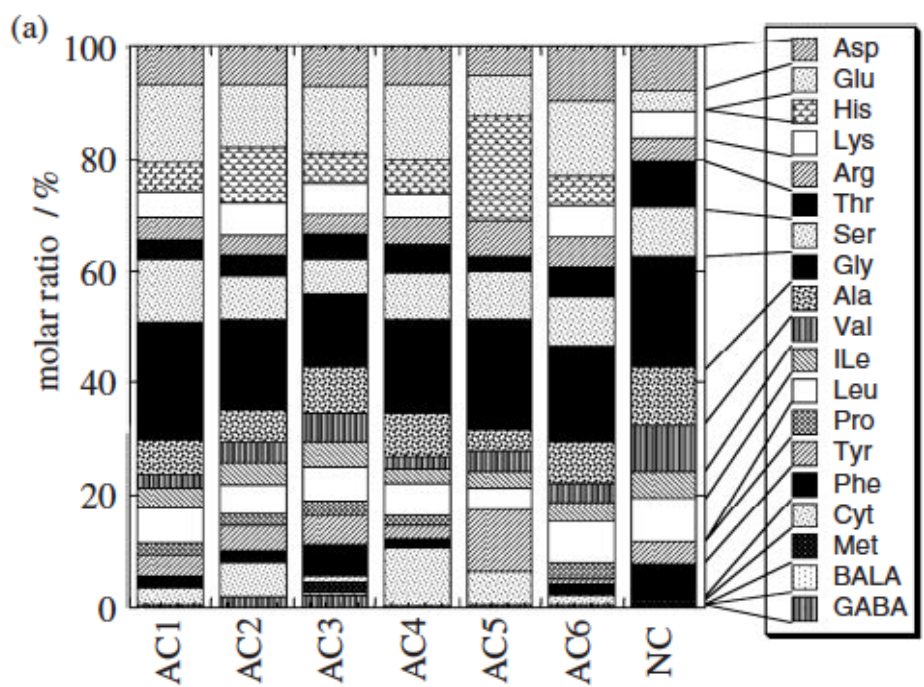

Figure 6

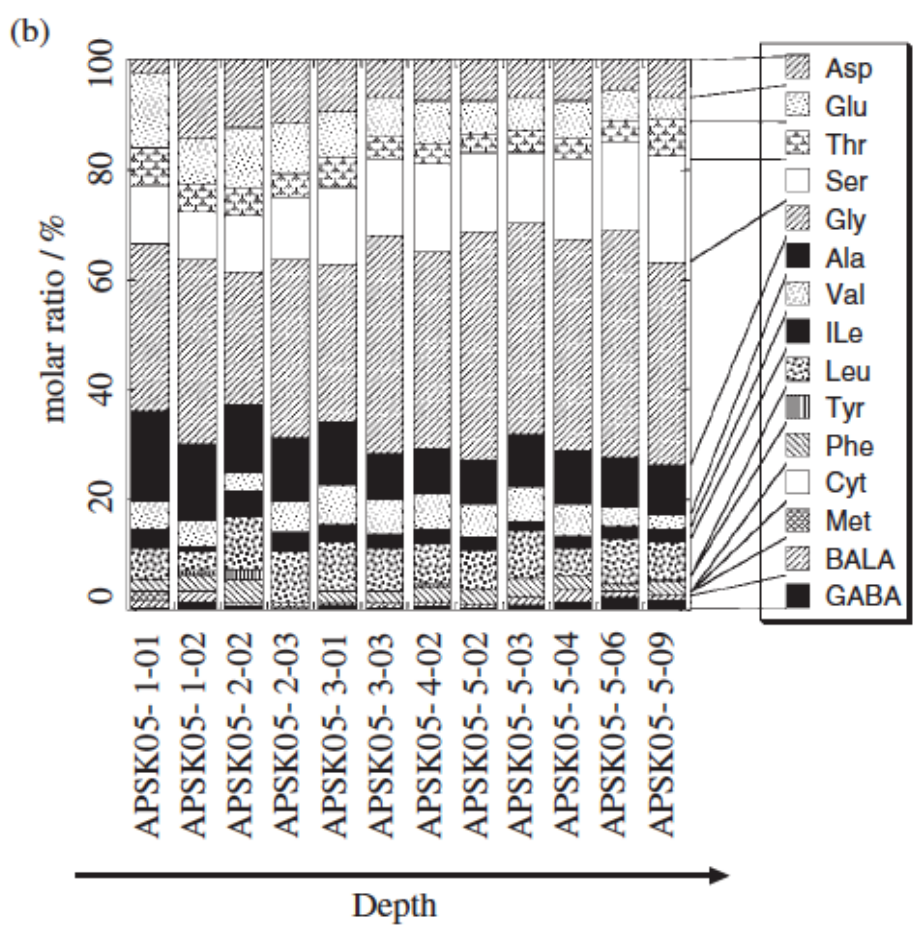

Figure 6

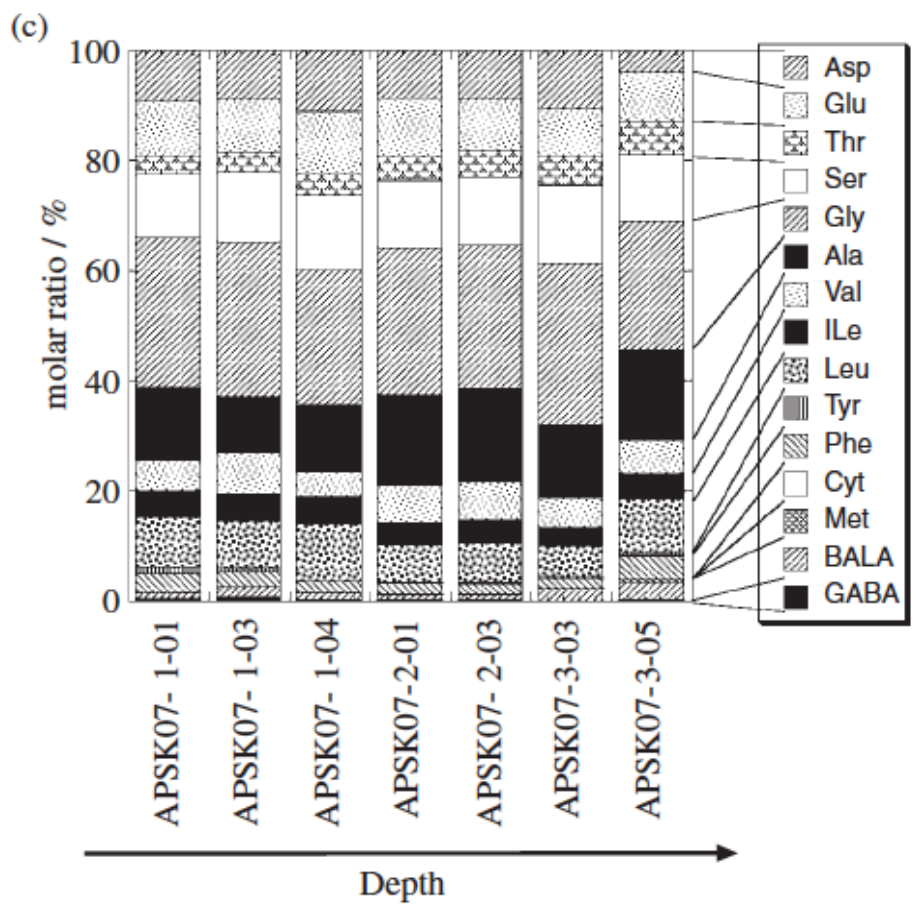

Fig. 6
Molar ratio of hydrolyzed amino acid in deep-sea hydrothermal chimney and core samples at sites APSK 05 and 07. Each marker indicates the individual amino acid component. AC stands for artificial chimney formed by the Kuwabara type *in situ* incubator system, and NC stands for natural chimney recovered by submersible investigation. Fundamental data of core samples are referred from the Ref. 19 and 20.

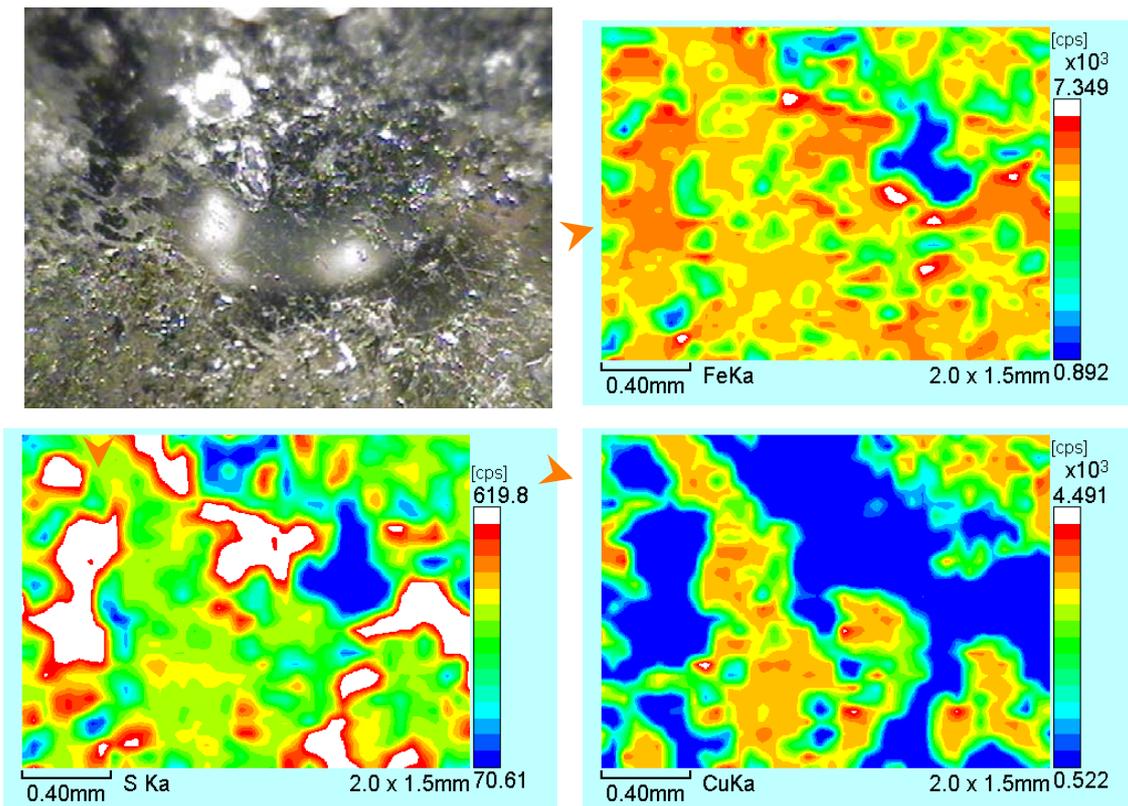

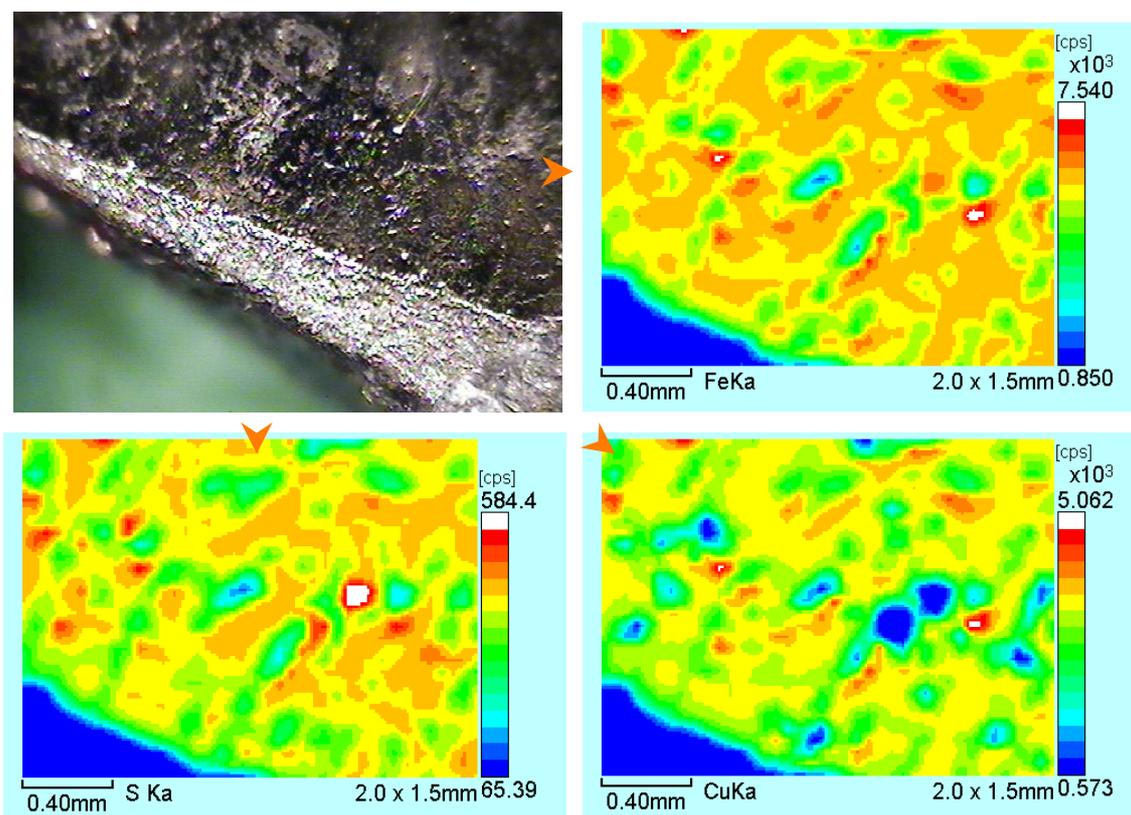

Figure 7

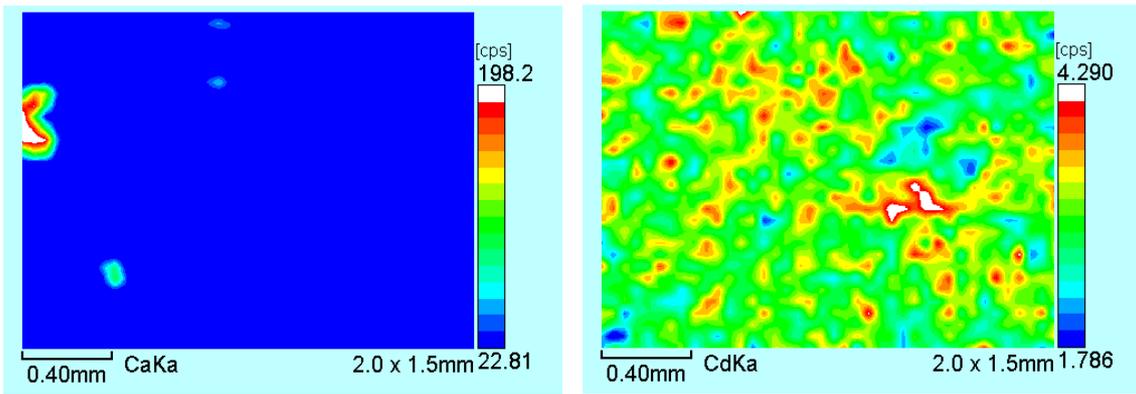

Fig. 7
Simultaneous elemental mapping image of natural chimney (NC) samples scanning by energy dispersive X-ray spectrometer (mEDX) for the natural chimney at Suiyo seamount, Izu-Bonin arc, Pacific ocean. We could not analyze AC samples since the trapped sulfide matrix were not solidified.

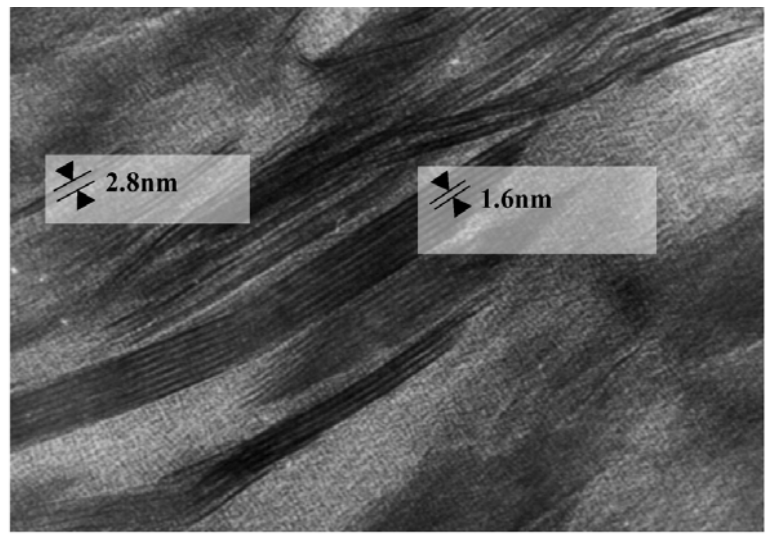

Figure 8

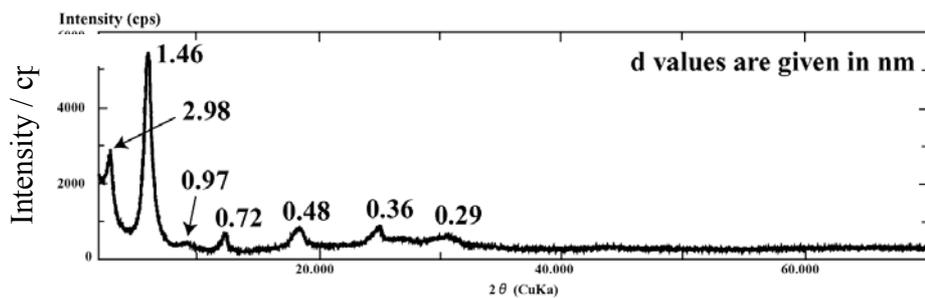

Fig. 8
TEM image and XRD chart of chlorite / montmorillonite from APSK 07-2-01.

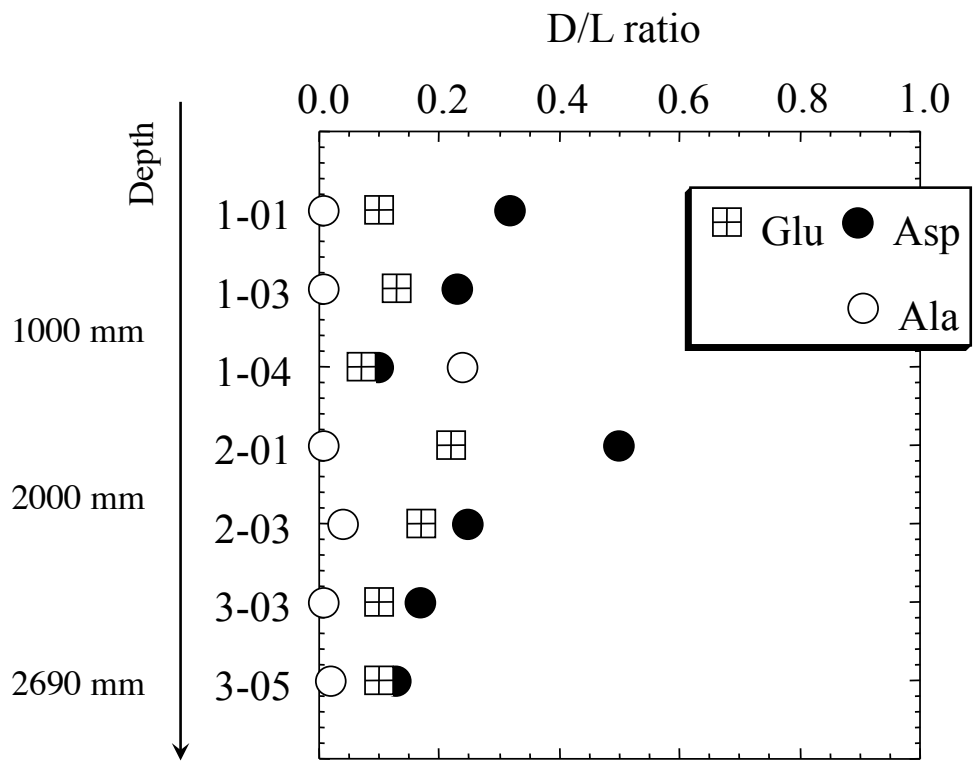

Fig. 9
Vertical distribution of chiral ratio (D/L ratio) of L-from and D-form amino acid enantiomers at APSK 07 site.

Table 1

Concentration of amino compounds trapped in artificial chimney (AC) and natural vent chimney (NC) in submarine hydrothermal systems at Suiyo seamount, Izu-Bonin arc, Pacific ocean. Sample name of NC was HY#12-CM. Those samples from AC1 to AC6 were collected by the same experiment using KI incubator.

| Characteristics | Amino acid | Sample number | | | | | | |
| --- | --- | --- | --- | --- | --- | --- | --- | --- |
| | | AC1 | AC2 | AC3 | AC4 | AC5 | AC6 | NC |
| | portion | interior | interior | interior | interior | interior | exterior | interior |
| Acidic | | | | | | | | |
| | Aspartic acid | 3.47 | 1.77 | 3.55 | 2.94 | 0.56 | 6.18 | 30.34 |
| | Glutamic acid | 6.78 | 2.80 | 5.72 | 5.79 | 0.75 | 8.55 | 14.18 |
| Basic | | | | | | | | |
| | Histidine | 2.61 | 2.51 | 2.68 | 2.64 | 2.00 | 3.56 | - |
| | Ornithine | n.d. | n.d. | n.d. | n.d. | n.d. | n.d. | - |
| | Lysine | 2.47 | 1.53 | 2.83 | 2.01 | n.d. | 3.43 | 18.32 |
| | Arginine | 1.86 | 0.86 | 1.60 | 2.10 | 0.68 | 3.39 | 15.36 |
| Neutral | | | | | | | | |
| Hydroxy | | | | | | | | |
| | Threonine | 1.79 | 0.96 | 2.28 | 2.20 | 0.27 | 3.38 | 32.12 |
| | Serine | 5.50 | 1.90 | 3.01 | 3.59 | 0.92 | 5.69 | 32.62 |
| Straight | | | | | | | | |
| | Glycine | 10.55 | 4.19 | 6.38 | 7.37 | 2.11 | 10.83 | 76.28 |
| | Alanine | 3.03 | 1.36 | 3.99 | 3.33 | 0.40 | 4.78 | 39.13 |
| Branched | | | | | | | | |
| | Valine | 1.21 | 0.97 | 2.57 | 0.92 | 0.39 | 2.22 | 31.24 |
| | Isoleucine | 1.81 | 0.98 | 2.10 | 1.20 | 0.29 | 2.26 | 18.18 |
| | Leucine | 3.19 | 1.38 | 3.07 | 2.45 | 0.44 | 4.68 | 30.05 |
| Secondary | | | | | | | | |
| | Proline | 1.01 | 0.55 | 1.28 | 0.85 | n.d. | 1.82 | - |
| Aromatic | | | | | | | | |
| | Tyrosine | 1.91 | 1.14 | 2.49 | 1.09 | 1.17 | 0.44 | 15.31 |
| | Phenylalanine | 1.07 | 0.54 | 2.84 | 0.61 | n.d. | 1.43 | 24.82 |
| Sulfur-containing | | | | | | | | |
| | Cysteine | 1.36 | 1.53 | 0.42 | 4.58 | 0.61 | 0.95 | n.d. |
| | Methionine | n.d. | n.d. | 0.88 | n.d. | n.d. | n.d. | 3.67 |
| Non-protein | | | | | | | | |
| | β-Alanine | n.d. | n.d. | 0.29 | n.d. | n.d. | n.d. | 0.50 |
| | γ-Aminobutyric acid | 0.32 | 0.46 | 1.07 | 0.07 | 0.07 | 0.41 | 0.50 |
| **Total hydrolyzed amino acids** | | 49.93 | 25.43 | 49.04 | 43.74 | 10.66 | 63.98 | 382.62 |
| Hexosamine | | | | | | | | |
| | Glucosamine | n.d. | n.d. | n.d. | n.d. | n.d. | 3.53 | 1.44 |
| | Galactosamine | n.d. | n.d. | n.d. | n.d. | n.d. | 4.56 | 1.60 |
| **Total hydrolyzed amino sugars** | | 0.00 | 0.00 | 0.00 | 0.00 | 0.00 | 8.09 | 3.03 |

Each value stands for nmol/g-rock. n.d.: not detected, tr.: trace amount (detected but not quantified)

Table 2

Molar ratio (mole%) of amino compounds trapped in artificial chimney (AC) and natural vent chimney (NC) in submarine hydrothermal systems at Suiyo seamount, Izu-Bonin arc, Pacific ocean.

| Mole % | Amino acid | Sample number | | | | | | NC |
|---|---|---|---|---|---|---|---|---|
| | | AC1 | AC2 | AC3 | AC4 | AC5 | AC6 | |
| Acidic | | | | | | | | |
| | Aspartic acid | 6.9 | 7.0 | 7.2 | 6.7 | 5.2 | 9.7 | 7.9 |
| | Glutamic acid | 13.6 | 11.0 | 11.7 | 13.2 | 7.1 | 13.4 | 3.7 |
| Basic | | | | | | | | |
| | Histidine | 5.2 | 9.9 | 5.5 | 6.0 | 18.8 | 5.6 | 0.0 |
| | Ornithine | 0.0 | 0.0 | 0.0 | 0.0 | 0.0 | 0.0 | 0.0 |
| | Lysine | 4.9 | 6.0 | 5.8 | 4.6 | 0.0 | 5.4 | 4.8 |
| | Arginine | 3.7 | 3.4 | 3.3 | 4.8 | 6.4 | 5.3 | 4.0 |
| Neutral | | | | | | | | |
| Hydroxy | | | | | | | | |
| | Threonine | 3.6 | 3.8 | 4.7 | 5.0 | 2.5 | 5.3 | 8.4 |
| | Serine | 11.0 | 7.5 | 6.1 | 8.2 | 8.6 | 8.9 | 8.5 |
| Straight | | | | | | | | |
| | Glycine | 21.1 | 16.5 | 13.0 | 16.8 | 19.8 | 16.9 | 19.9 |
| | Alanine | 6.1 | 5.4 | 8.1 | 7.6 | 3.7 | 7.5 | 10.2 |
| Branched | | | | | | | | |
| | Valine | 2.4 | 3.8 | 5.2 | 2.1 | 3.7 | 3.5 | 8.2 |
| | Isoleucine | 3.6 | 3.8 | 4.3 | 2.7 | 2.7 | 3.5 | 4.8 |
| | Leucine | 6.4 | 5.4 | 6.3 | 5.6 | 4.1 | 7.3 | 7.9 |
| Secondary | | | | | | | | |
| | Proline | 2.0 | 2.2 | 2.6 | 1.9 | 0.0 | 2.8 | 0.0 |
| Aromatic | | | | | | | | |
| | Tyrosine | 3.8 | 4.5 | 5.1 | 2.5 | 11.0 | 0.7 | 4.0 |
| | Phenylalanine | 2.1 | 2.1 | 5.8 | 1.4 | 0.0 | 2.2 | 6.5 |
| Sulfur-containing | | | | | | | | |
| | Cysteine | 2.7 | 6.0 | 0.9 | 10.5 | 5.7 | 1.5 | 0.0 |
| | Methionine | 0.0 | 0.0 | 1.8 | 0.0 | 0.0 | 0.0 | 1.0 |
| Non-protein | | | | | | | | |
| | β-Alanine | 0.0 | 0.0 | 0.6 | 0.0 | 0.0 | 0.0 | 0.1 |
| | γ-Aminobutyric acid | 0.6 | 1.8 | 2.2 | 0.2 | 0.7 | 0.6 | 0.1 |
| | **Total** | 100 | 100 | 100 | 100 | 100 | 100 | 100 |
| Hexosamine | | | | | | | | |
| | Glucosamine | - | - | - | - | - | 43.7 | 47.4 |
| | Galactosamine | - | - | - | - | - | 56.3 | 52.8 |
| | **Total** | - | - | - | - | - | 100 | 100 |

Table 3

Concentration and D/L ratios of amino acids in submarine hydrothermal sub-vent core samples of APSK 07 at Suiyo seamount, Izu-bonin arc, Pacific ocean (cf. APSK05 in the Ref. 20.)

| Characteristics | Amino acid | Sample number | | | | | | |
|---|---|---|---|---|---|---|---|---|
| | | 1-01 | 1-03 | 1-04 | 2-01 | 2-03 | 3-03 | 3-05 |
| Acidic | | | | | | | | |
| | Aspartic acid | 8.04 | 3.07 | 4.44 | 5.84 | 4.63 | 3.80 | 1.17 |
| | Glutamic acid | 8.62 | 3.29 | 4.75 | 7.15 | 4.74 | 3.15 | 2.57 |
| Neutral | | | | | | | | |
| Hydroxy | | | | | | | | |
| | Threonine | 3.00 | 1.31 | 1.52 | 2.85 | 2.60 | 1.89 | 1.76 |
| | Serine | 9.81 | 4.36 | 5.44 | 8.23 | 6.33 | 5.22 | 3.52 |
| Straight | | | | | | | | |
| | Glycine | 23.65 | 9.73 | 9.87 | 17.80 | 13.48 | 10.58 | 6.81 |
| | Alanine | 11.26 | 3.57 | 4.99 | 11.01 | 8.70 | 4.79 | 4.73 |
| Branched | | | | | | | | |
| | Valine | 4.94 | 2.52 | 1.79 | 4.57 | 3.55 | 1.90 | 1.78 |
| | Isoleucine | 3.94 | 1.76 | 2.02 | 2.60 | 2.19 | 1.22 | 1.38 |
| | Leucine | 7.83 | 2.92 | 4.13 | 4.55 | 3.79 | 2.14 | 2.86 |
| Aromatic | | | | | | | | |
| | Tyrosine | 1.08 | 0.27 | tr. | 0.12 | 0.11 | 0.05 | 0.14 |
| | Phenylalanine | 2.98 | 0.88 | 0.91 | 1.29 | 0.94 | 0.66 | 1.26 |
| Sulfur-containing | | | | | | | | |
| | Cysteine | n.d. | n.d. | n.d. | n.d. | tr. | tr. | n.d. |
| | Methionine | 0.20 | tr. | tr. | 0.21 | 0.12 | tr. | 0.08 |
| Non-protein | | | | | | | | |
| | β-Alanine | 0.77 | 0.65 | 0.54 | 0.60 | 0.47 | 0.83 | 0.96 |
| | α-Aminobutyric acid | 0.11 | n.d. | n.d. | n.d. | n.d. | n.d. | n.d. |
| | γ-Aminobutyric acid | 0.50 | 0.31 | 0.10 | 0.17 | 0.10 | tr. | 0.08 |
| | δ-Aminovaleic acid | 0.74 | tr. | tr. | tr. | tr. | tr. | tr. |
| | α-Aminoadipic acid | 0.10 | 0.12 | 0.09 | 0.15 | tr. | 0.25 | tr. |
| **Total hydrolyzed amino acids** | | 87.58 | 34.75 | 40.59 | 67.14 | 51.74 | 36.48 | 29.11 |

Each value stands for nmol/g-rock.  n.d.: not detected, tr.: trace amount (detected but not quantified)

| D/L ratio | Aspartic acid | 0.32 | 0.23 | 0.10 | 0.50 | 0.25 | 0.17 | 0.13 |
|---|---|---|---|---|---|---|---|---|
| | Glutamic acid | 0.10 | 0.13 | 0.07 | 0.22 | 0.17 | 0.10 | 0.10 |
| | Alanine | 0.01 | 0.01 | 0.24 | 0.01 | 0.04 | 0.01 | 0.02 |

Each value stands for D- and L- stereo isomer ratio.